\documentclass[preprint,12pt]{elsarticle}

\usepackage{graphicx}%
\usepackage{multirow}%
\usepackage{amsmath,amssymb,amsfonts}%
\usepackage{amsthm}%
\usepackage{mathrsfs}%
\usepackage[title]{appendix}%
\usepackage{xcolor, soul}%
\usepackage{textcomp}%
\usepackage{manyfoot}%
\usepackage{booktabs}%
\usepackage{algorithm}%
\usepackage{algorithmicx}%
\usepackage{algpseudocode}%
\usepackage{listings}%
\usepackage{verbatim}
\usepackage{subcaption}
\usepackage{graphicx}
\usepackage{physics}
\usepackage{url}
\usepackage{units}
\usepackage{natbib}
\usepackage{ulem}

\def\etal.{et\penalty50\ al.}
\graphicspath{ {figures/} }

\journal{Computers in Biology and Medicine}

\newcommand{\revs}[1]{{#1}} 
\newcommand{\reva}[1]{{#1}} 
\newcommand{\revb}[1]{{#1}} 

\newcommand{\revss}[1]{{#1}} 
\newcommand{\revaa}[1]{{#1}} 
\newcommand{\revbb}[1]{{#1}} 
\newcommand{\revbbrem}[1]{} 

\begin{document}

\begin{frontmatter}

\title{Patient-specific coronary angioplasty simulations -- a mixed-dimensional finite element modeling approach}

\author[1,2]{Janina C. Datz\corref{cor1}}\ead{janina.datz@tum.de}
\author[3]{Ivo Steinbrecher}
\author[1]{Christoph Meier}
\author[3]{Nora Hagmeyer}
\author[2]{Leif-Christopher Engel}
\author[3]{Alexander Popp}
\author[4]{Martin R. Pfaller}
\author[2]{Heribert Schunkert}
\author[1,5]{Wolfgang A. Wall}

\cortext[cor1]{Corresponding author}

\affiliation[1]{organization={Institute for Computational Mechanics, Technical University of Munich},
            country={Germany}}

\affiliation[2]{organization={Department of Cardiology, Deutsches Herzzentrum München, Technical University of Munich},
            country={Germany}}

\affiliation[3]{organization={Institute for Mathematics and Computer-Based Simulation, University of the Bundeswehr Munich},
            country={Germany}}
\affiliation[4]{organization={Pediatric Cardiology, Cardiovascular Institute, and Institute for Computational and Mathematical Engineering, Stanford University},
            country={USA}}
\affiliation[5]{organization={Munich Institute of Biomedical Engineering, Technical University of Munich},
            country={Germany}}

\begin{abstract}

Coronary angioplasty with stent implantation is the most frequently used interventional treatment for coronary artery disease.
However, reocclusion within the stent, referred to as in-stent restenosis, occurs in up to 10\% of lesions.
It is widely accepted that mechanical loads on the vessel wall strongly affect adaptive and maladaptive mechanisms.  
Yet, the role of procedural and lesion-specific influence on restenosis risk remains understudied.
Computational modeling of the stenting procedure can provide new mechanistic insights, such as local stresses, that play a significant role in tissue growth and remodeling.
Previous simulation studies often featured simplified artery and stent geometries and cannot be applied to real-world examples. 
Realistic simulations were computationally expensive since they featured fully resolved stenting device models. 
The aim of this work is to develop and present a mixed-dimensional formulation to simulate the patient-specific stenting procedure with a reduced-dimensional beam model for the stent and 3D models for the artery. 
In addition to presenting the numerical approach, we apply it to realistic cases to study the intervention's mechanical effect on the artery and correlate the findings with potential high-risk locations for in-stent restenosis. 
We found that high artery wall stresses develop during the coronary intervention in severely stenosed areas and at the stent boundaries. 
Herewith, we lay the groundwork for further studies towards preventing in-stent restenosis after coronary angioplasty.  

\end{abstract}




\end{frontmatter}

\section{Introduction}
\label{sec:introduction}

Coronary artery disease (CAD) restricts the supply of oxygenated blood through the coronary arteries to the myocardial tissue. 
It commonly stems from atherosclerosis and can lead to ischemia or even myocardial infarction~\cite{timmis_european_2022}. 
Worldwide, CAD accounts for over 16\% of total deaths and is therefore amongst the leading causes of morbidity and mortality~\cite{noauthor_gbd_nodate}. 
If cardiologists diagnose CAD with one or more severe obstructions, the affected vessels are usually reopened using percutaneous coronary intervention (PCI), also known as coronary angioplasty.
This procedure often includes coronary stenting in addition to balloon angioplasty.
The intervention involves introducing a guidewire to the stenotic site, guided by coronary angiography imaging. 
A balloon, which is folded on a catheter, is inflated to open the occlusion.
A stent is implanted to provide radial support, preventing elastic recoil~\cite{ahmad_percutaneous_2023}.
However, repeated revascularization is necessary for patients presenting with symptoms because of in-stent restenosis (ISR), usually 3-12 months after the PCI~\cite{burzotta_percutaneous_2018}. 
Even though new generations of drug-eluting stents are used, ISR occurs in about 10\% of stented lesions~\cite{byrne_stent_2015}.
Given the high global incidence of CAD, this number is still significant.
While ISR is a multifactorial disease whose causes are not yet fully understood, it is widely accepted that biological factors, such as neointimal hyperplasia and neoatherosclerosis formation, but also other factors, like stent underexpansion or malapposition play essential roles~\cite{byrne_stent_2015, shafiabadihassani_InStentRestenosisOverview_2024}.
It is also obvious that the continuous exposure to mechanical loads after the stent implantation leads to changes in the mechanobiological equilibrium, which causes cell-mediated mechanoregulation in soft tissue, involving long-term growth and remodeling processes~\cite{cyron_VascularHomeostasisConcept_2014, cyron_HomogenizedConstrainedMixture_2016}.
Additionally, subintimal collagen damage due to overstretch in the media during the stent implantation promotes inflammatory processes, which interact with physiological mechanobiological progressions~\cite{gierig_PostangioplastyRemodelingCoronary_2024, marino_MolecularlevelCollagenDamage_2019}. 
Endothelial denudation and disturbances in the blood-flow-related wall shear stress can additionally advance inflammation~\cite{escuer_MathematicalModellingRestenosis_2019, gierig_PostangioplastyRemodelingCoronary_2024}.
Local mechanical factors during and after the stent implantation highly influence these mechanisms.
However, clinicians are restricted to relying on statistically determined patient-, stent-type- or procedure-related risk factors, such as comorbidities, the affected artery segment, or stent type and length~\cite{cassese_incidence_2014, guldener_machine_2023,condello_stent_2023, rohman_coronary_2023}. 
Information on the local wall stresses that are known to strongly affect growth and remodeling processes, in correlation to the individual lesion could help deepen the understanding of personalized risk assessment for ISR.


Physics-based computational modeling and simulation provide insight into these mechanical factors, which cannot be directly measured or retrieved from imaging modalities. 
Recent reviews highlight the role of biomechanical factors, such as shear stress and intramural stress, in the onset and progression of ISR~\cite{jenei_WallShearStress_2016, antonini_ReviewUseFinite_2023}. 
To gain insight into the biomechanical state of an individual artery during PCI, a computational model of the respective vessel is necessary, as well as a realistic mathematical model of the external loads caused by balloon inflation and stent deployment. 
The simulation of coronary angioplasty and stenting procedures has been the subject of various studies. 
Elaborate reviews of the methods and objectives of significant contributions in this field can be found e.g. in~\cite{holzapfel_ComputationalApproachesAnalyzing_2014, chiastra_modeling_2021, martin_computational_2011}. 
Computational studies regarding the progression of ISR can be divided into structural modeling, computational fluid dynamics studies, and simulation of drug diffusion into the tissue~\cite{morlacchi_modeling_2013}. 
This work focuses on the structural modeling of PCI.

The challenges in structural PCI modeling can be attributed to the three main components, i.e., artery, balloon, and stent.
The artery model was often represented with simplified generic geometries since the specific geometry was not the main subject of coronary angioplasty simulation studies~\cite{holzapfel_simulation_2006, conway_computational_2012, rahinj_numerical_2021, mortier_virtual_2011}. 
Generic artery models were also used to study microscopic processes related to ISR~\cite{holzapfel_simulation_2006, boyle_application_2013, pericevic_influence_2009, manjunatha_FiniteElementModelling_2022, cheng_finite_2021}.
Using short, straight artery sections or circular cross-sections reduces the modeling complexity and allows for the inspection of specific objectives. 
In this study, however, we focus on the mechanical effect of PCI on realistic individual arteries. Therefore, the developed approach must be applicable to patient-specific coronary artery models.
The artery geometry significantly influences the stent deployment and the stresses and strains in the stented area. 
In disease, the wall thickness shows large variations, leading to locally varying wall properties.
Also, typical diseased artery characteristics, such as eccentric plaques~\cite{jiang_numerical_2019} and highly tortuous arteries~\cite{du_value_2023}, are known to influence PCI success.
Simulations of the fully resolved stent deployment process with patient-specific coronary artery models have been demonstrated e.g. in~\cite{mortier_novel_2010, morlacchi_patient-specific_2013, samant_computational_2021, chiastra_computational_2016, zahedmanesh_simulation_2010, ragkousis_computational_2015,antonini_ComputationalWorkflowModeling_2024}. 
Detailed balloon unfolding and stent expansion modeling in these studies provided valuable insights into the stenting mechanics.
However, these approaches are computationally very expensive and hence difficult to re-use for larger studies.
For example, such a setup requires relatively fine 3D stent and balloon discretizations, often exceeding the number of elements used for the artery.
The modeling of the stent expansion through the inflation of a coronary balloon catheter is complex, even without considering the interaction with a coronary artery~\cite{martin_computational_2011}. 
Balloon catheters used for stent implantation achieve their final, inflated size primarily by unfolding from their initial configuration.
Therefore, the expansion behavior differs from a comparable, tube-like isotropic elastic material when inflated~\cite{geith_experimental_2020, geith_importance_2019}. 
Simplified modeling approaches for the balloon have been proposed in~\cite{geith_importance_2019,ragkousis_simulation_2014,de_beule_realistic_2008}. 
They investigated modeling the balloon geometry as displacement-controlled or homogeneous, elastically deformable cylinders. 
This simplification allowed a more efficient simulation without the computationally demanding unfolding mechanism.
An improved approach to replicate the typical balloon inflation behavior using a customized orthotropic material formulation for elastically deforming balloon geometries has been presented by~\cite{kiousis_NumericalModelStudy_2007, kiousis_experimental_2009}. 
They tested the pressure-based expansion behavior on a straight stenting device setup and compared it to experimental balloon inflation data, delivering conforming results.
The pressure-based balloon inflation allows for a more realistic expansion behavior than prescribing the displacements.
In addition, this approach does not require the modeling of the complex balloon unfolding mechanics while still showing the radial inflation behavior typical for a folded balloon. 
The material properties and geometrical shape of the stent significantly dictate its ultimate form and expansion behavior, thereby imposing a substantial demand on the modeling of the stent to achieve accurate simulation results.
Several computational simulation studies focused on the stent's mechanical properties, such as their elastic recoil, radial support, and susceptibility to fracture or fatigue ~\cite{bobel_computational_2015, kumar_design_2019, migliavacca_predictive_2005, auricchio_FatigueMetallicStents_2016}.
Due to the slenderness of the stent struts, 3D-resolved stents require a very fine mesh and, hence, are computationally very expensive.
\revb{Reduced-dimensional (1D) beam finite elements provide an accurate and effective way to discretize slender geometries at a fraction of the degrees of freedom~\cite{zunino_integrated_2016}. }
\revb{Stents were modeled with 1D beam finite elements in~\cite{krewcun_fast_2019, hall_comparison_2006, tambaca_NovelApproachModeling_2010,canic_geometric_2023,tambaca_OnedimensionalQuasistaticModel_2015}, significantly reducing computational effort while retaining a realistic stent behavior.
However, in these studies, only the stent was modeled without interaction with the artery. 
Stent-artery-interaction in the context of drug-elusion from DES was modeled using 1D stents in~\cite{hossainy_MathematicalModelPredicting_2008, dangelo_ModelReductionStrategies_2011, cutri_DrugDeliveryPatterns_2013}, neglecting the previous stenting procedure or using a fully resolved system for this part. 
Mixed-dimensional interaction during stenting procedures was modeled using rigid endothelial surfaces~\cite{bisighini_EndoBeamsJlJulia_2022, bisighini_MachineLearningReduced_2023}, or embedding the stent into a graft~\cite{capelli_finite_2012}.
}
In~\cite{zunino_integrated_2016}, a comprehensive review of reduced-dimensional models for drug-eluting stents and their accuracy and performance compared to the 3D-resolved problem was provided.

In this work, we present a novel approach for the simulation of patient-specific coronary angioplasty,
including all components described above, i.e., artery, balloon, and stent.
One of the main novelties is the usage of 1D beam finite elements for the stent struts while still accurately modeling the interaction between the system components of different dimensionality (1D stent struts, 3D artery and balloon), requiring a mixed-dimensional coupling approach.
\revbb{This approach not only allows for a significant reduction of the degrees of freedom to discretize the stent, but also offers further advantages and efficiency gains and, hence, reduces the computational effort compared to a 3D stent discretization.}
We build our framework upon the methods introduced in~\cite{steinbrecher_mortar-type_2020,steinbrecher_consistent_2022,steinbrecher_consistent_2022-1}.
There, collocation and mortar-type methods are presented to consistently model mixed-dimensional interaction problems from embedded fibers to beam-solid contact.
We use a mortar finite element framework for the contact mechanics between the solid parts, as introduced in~\cite{popp_FiniteDeformationMortar_2009,popp_dual_2010}.
Furthermore, we employ a novel elasto-plastic constitutive model crafted for 1D beam finite elements to accurately represent the stent behavior in the post-elastic regime.
There, the elasto-plastic relations are formulated based on the beam cross-section stress and strain resultants. 
This results in a significant reduction in the computational and modeling effort for the stent while still accurately capturing local and global effects of the stent on the mechanical response of the artery.
Moreover, the employment of 1D beam finite elements for the stent struts allows for efficient and flexible modeling of the stenting device setup, which can be easily adjusted to accommodate different geometries.
For the balloon modeling, we reduce the simulation complexity by using a simplified geometry, which deforms elastically upon inflation. 
For this simplification, our approach is based on the methods in~\cite{kiousis_NumericalModelStudy_2007}, namely employing an artificially orthotropic balloon material formulation to achieve a realistic expansion behavior. 
For curved initial configurations, as in most patient-specific applications, the anisotropic material additionally prevents the balloon from extensive bending during the inflation.
This material formulation allows us to model the realistic balloon inflation behavior while using a simplified, easy-to-implement, and computationally efficient cylindrical balloon geometry in complex, patient-specific simulations. 

In summary, we present a novel approach specifically tailored for studying the effect of PCI on individual coronary arteries\revbb{, but also easily adaptable for other biomechanical problems involving stents}.
We believe the results presented in this work show that this can become a viable tool for incorporating image-derived, patient-specific geometries to facilitate objective comparisons of different lesions or treatment procedures. 
\revbb{Through the efficient mixed-dimensional approach, this work serves as an important step towards the broader applicability of computational PCI simulations for larger patient cohorts and clinical studies.}
\revb{In addition to the factors considered in the traditional ISR risk stratification strategies, insights from such comparisons could} inform the clinical assessment of lesions considered for stent implantation in the future.

\section{Computational models and methods} 
\label{sec:methods}

We formulate all components of the stenting device and the coronary artery in a geometrically and constitutively nonlinear framework with contact interface conditions between the bodies. 
Throughout this work, the term stenting device describes the coronary balloon and balloon-inflatable stent system. 
The artery and the balloon are 3D Boltzmann continua following a classical continuum solid mechanics approach, cf.~\cite{holzapfel_nonlinear_2002}. 
We employ a stent model based on reduced-dimensional, 1D beam elements following the Simo-Reissner theory~\cite{meier_geometrically_2019} with an elasto-plastic material model. 
The interaction between the balloon catheter and artery is modeled with computational contact mechanics using mortar methods; for the stent, we utilize a beam-to-solid surface contact approach~\cite{steinbrecher_consistent_2022}.
Spatial discretization of all continua is based on finite elements and all simulations are performed with our in-house multiphysics high-performance code 4C~\cite{wall_baci_2018}.

\subsection{Coronary artery model}
\label{sec:arterymodel}

\begin{figure}[h]
    \centering
    \includegraphics[width=\textwidth]{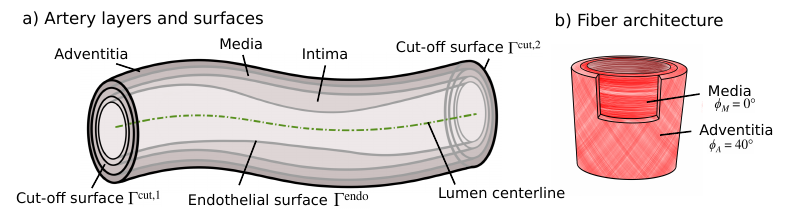}
    \caption{Artery macro- and microstructure}
    \label{fig:arteryfibers}
\end{figure}

The artery material is modeled as being nearly incompressible, anisotropic, and hyperelastic.
We introduce the virtual work $\delta W$ formulated in the reference configuration as the basis for the finite element method. 
Inertial forces are neglected for the mechanical equilibrium as we are not interested in any dynamic process. 
Hence, we will later use a quasi-static pseudo-time discretization and apply the loads in small load steps. 
We derive the boundary value problem from the balance equations and constitutive relations. 
Together with the boundary conditions for the given solid body, it reads 
\begin{equation}
    \label{eq:artery_bvp}
    \delta W = \int_{\Omega_{0}^{\mathcal{S}}} \boldsymbol{S}:\delta\boldsymbol{E}~\mathrm{d}V + \int_{\Gamma_{0}^{\mathrm{cut}}} c_{c} \boldsymbol{u} \cdot \delta \boldsymbol{u}~\mathrm{d}A + \int_{\Gamma_{0}^{\mathrm{endo}}} p_{d} \boldsymbol{F}^{-T} \boldsymbol{N}_{0} \cdot \delta \boldsymbol{u}~\mathrm{d}A = 0\; .
\end{equation}
Here, $\boldsymbol{u}$ is the displacement field, $\boldsymbol{F}$ denotes the deformation gradient, $\boldsymbol{S}$ the second Piola-Kirchoff stress tensor, $\boldsymbol{E}$ is the Green-Lagrange strain and $\boldsymbol{N}_{0}$ is the reference surface normal direction (further details can be found in~\ref{sec:app_solidmechanics})~\cite{holzapfel_nonlinear_2002}.
The integral over the cut-off surfaces $\Gamma_{0}^{\mathrm{cut}} = \Gamma_{0}^{\mathrm{cut,1}} + \Gamma_{0}^{\mathrm{cut,2}}$ describes the boundary condition on the artery end surfaces, see Figure~\ref{fig:arteryfibers}. 
In the literature, these surfaces were often fixed with displacement boundary conditions, cf.~\cite{holzapfel_layer-specific_2002,gijsen_simulation_2008}. 
However, this approach introduces strong boundary effects and hence requires an extended artery model length.
In this work, we use omnidirectional springs as a reasonable approximation for realistic boundary conditions while reducing boundary effects.
Coronary arteries are generally prestressed in their initial configuration, which influences their biomechanical response to the stent implantation and the overall stress state~\cite{ramella_NecessityIncludeArterial_2023}. 
Since this work does not focus on the exact initial geometry, we apply the blood pressure $p_{d} = \unit[85]{mmHg}$ directly to the endothelial surface $\Gamma_{0}^{\mathrm{endo}}$, which leads to small deformations compared to the deformations during the stenting procedure.
The blood pressure is averaged over time since the time frame of the simulation is small compared to its periodic changes. 
The set of nonlinear partial differential equations of the boundary value problem is discretized in space using the finite element method with second-order tetrahedral elements. 

The artery wall consists of three layers, the intima, media, and adventitia, which have different material properties and microstructures~\cite{gasser_hyperelastic_2005}, see Figure~\ref{fig:arteryfibers}a).
The media absorbs most of the stress in the physiological range, while the adventitia serves as a protective layer against overstretch and rupture at high stresses. 
Each layer's ground material consists of elastin fibers, proteoglycans, and, in the media, smooth muscle cells~\cite{holzapfel_biomechanics_2017}. 
The layered walls are strongly anisotropic because of their collagen component, which is significantly stiffer than its surrounding material.
The constitutive description of soft biological tissues on a macroscopic level must, therefore, include its fibrous structure and the resulting preferred directions associated with the fibers~\cite{cowin_new_2004, holzapfel_biomechanics_2017}. 
We use an exponential material for the fiber contributions, which are embedded in a Neo-Hookean base material. 
We add a material dissipation term with a viscoelastic pseudo-potential for numerical stability; the term is small compared to the elastic contributions. 
Details on the specific strain energies are given in~\ref{sec:app_strainenergies}.
The artery material constants in this work are chosen similarly to~\cite{gasser_finite_2007} based on the experimental work in~\cite{holzapfel_determination_2005}.
The Neo-Hookean material model for the isotropic part depends on the Poisson's ratio $\nu = 0.45$ and the Young's modulus $E$, which is $\unit[156.6]{kPa}$ and $\unit[15.66]{kPa}$ for the media and the adventitia, respectively. 
The material model for the elastic fiber contribution depends on the exponential constants $k_1$ and $k_2$~\cite{cowin_new_2004}, which exhibit exponential stiffening at higher deformations. 
The anisotropic material constants $k_1$ and $k_2$ are $\unit[0.64]{kPa}$ and $3.54$ for the media and $\unit[5.1]{kPa}$ and $15.4$ for the adventitia. 
The collagen fiber architecture differs in each layer~\cite{holzapfel_determination_2005} and is displayed for a generic artery extract in Figure~\ref{fig:arteryfibers}b).
Experimental research determined the media's collagen fiber architecture to be almost circumferential, i.e., $\phi_M=0^\circ$~\cite{holzapfel_determination_2005}. 
The adventitia's fibers are oriented approximately $\phi_A=\pm 40^\circ$ from the circumferential direction, forming a double helix structure with two fiber families in opposite directions.

\subsection{Stent model}
\label{sec:stentmodel}

\begin{figure}[h]
    \centering
    \includegraphics[width=\textwidth]{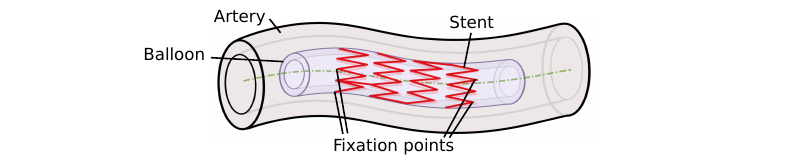}
    \caption{Setup of artery, balloon, and stent; the fixation points are only free to move in the radial direction.}
    \label{fig:arterysetup}
\end{figure}

We model the stent structure using 1D beam elements with an elasto-plastic constitutive law.
Figure~\ref{fig:arterysetup} shows the stent setup and fixation points, i.e., the nodes for the application of the Dirichlet boundary conditions, only allowing radial displacement.
If the stent centerline is straight, one additional stent node is fixed against axial movement to prevent slippage from the balloon since the contact is frictionless. 

\begin{figure}[h]
    \centering
    \includegraphics[width=\textwidth]{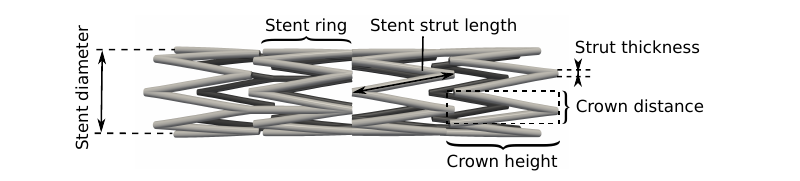}
    \caption{Stent geometry; the stent struts are represented with their true thickness, though modeled as 1D beam elements}
    \label{fig:stentstructure}
\end{figure}

The stent geometry follows a ``corrugated rings'' design with straight connectors, comparable to BioMatrix stent designs~\cite{biomatrix_stent}, see Figure~\ref{fig:stentstructure}. 
The geometry is constructed using the open-source 3D beam generator MeshPy~\cite{steinbrecher_meshpy_2023}.
The stent model consists of rings with eight crowns in each ring; the number of rings determines the stent length.
The distance between the rings is $\unit[1.5]{mm}$; the crown distance is $\unit[0.45]{mm}$ and the crown height is $\unit[1.5]{mm}$. 
The stent strut thickness is $\unit[0.12]{mm}$, and its cross-section is rectangular with rounded corners. 

Considering the slender shape of the stent struts, a reduced-dimensional 1D beam theory is a good approximation.
Each stent strut is discretized with three beam finite elements. 
The beams are modeled as 1D Cosserat continua embedded in 3D space based on the geometrically exact Simo-Reissner beam theory~\cite{reissner_one-dimensional_1972,simo_dynamics_1988}. 
\revb{We model the connection points between the beam elements as rigid joints, assuming that the flexibility of the joint itself can be neglected, which is a valid assumption for typical joints of 1D beam elements~\cite{klarmann_Coupling2DContinuum_2022}.}
The beam elements are described by the centerline curve $\boldsymbol{r}(s) \in \mathbb{R}^3$ connecting the cross-section centroids, where $s \in [0,L] \subset \mathbb{R}$ is an arc-length coordinate and $L$ the length of the beam in the undeformed configuration. 
The orientations of the beam cross-sections are described by the rotation tensors $\boldsymbol{\Lambda}(s) \in SO(3)$, where $SO(3)$ represents the rotation group. 
The corresponding variations of these kinematic quantities,~i.e., the virtual displacements and rotations, are denoted as $\delta \boldsymbol{r}$ and $\boldsymbol{\theta}$, respectively. 
Eventually, the weak form of the mechanical balance equations is given as
\begin{align}
\label{weakformspatial}
\!\!\!\!\!\! \delta W =  \int \limits_0^L \bigg( \delta \boldsymbol{\Omega}^T  \boldsymbol{M}^\mathcal{B}  +  \delta \boldsymbol{\Gamma}^T\boldsymbol{F}^\mathcal{B} - \delta \boldsymbol{\theta}^T  \tilde{\boldsymbol{m}} - \delta \boldsymbol{r}^T \tilde{\boldsymbol{f}} \bigg) ds
-\Big[\delta \boldsymbol{r}^T  \boldsymbol{f} \Big]_{\varGamma_{\sigma}} -  \Big[\delta \boldsymbol{\theta}^T  \boldsymbol{m} \Big]_{\varGamma_{\sigma}} = \,0.\!\!\!\!\!\!
\end{align}
Here, $\tilde{\boldsymbol{f}}$ and $\tilde{\boldsymbol{m}}$ correspond to externally applied distributed forces and moments, respectively, and $\varGamma_{\sigma}$ refers to the (Neumann) boundary of the beam with prescribed external forces and moments. 
In this context, $\boldsymbol{F}^\mathcal{B}$ and $\boldsymbol{M}^\mathcal{B}$ represent the (convected) force and moment stress resultants acting in the beam cross-sections. 
Note that we use the superscript $(.)^\mathcal{B}$ for ``beams'' here to avoid confusion with the deformation gradient of the solid parts.
Moreover, $\boldsymbol{\Gamma}=\boldsymbol{\Lambda}^T\!\boldsymbol{r}^{\prime}  \!-\! \boldsymbol{E}_1$ with $\boldsymbol{E}_1=[1,0,0]^T$ and $\boldsymbol{S}(\boldsymbol{\Omega}) \!=\! \boldsymbol{\Lambda}^T \!\boldsymbol{\Lambda}^{\prime} \!-\! \boldsymbol{\Lambda}_0^T\! \boldsymbol{\Lambda}_0^{\prime}$ are (convected) deformation measures that are work-conjugated to the stress resultants $\boldsymbol{F}^\mathcal{B}$ and $\boldsymbol{M}^\mathcal{B}$. 
Here, the notation $\boldsymbol{S}(\boldsymbol{\Omega})$ refers to the skew-symmetric tensor associated with the axial vector $\boldsymbol{\Omega}$. 
The prime $(.)' = \partial (.) / \partial s$ denotes the derivative with respect to the arc-length coordinate $s$, and the subscript $(.)_0$ refers to the (stress-free) undeformed configuration of the beam. 
In particular, the vector $\boldsymbol{\Gamma}$ represents axial tension and shear (two components), while the vector $\boldsymbol{\Omega}$  represents torsion and bending (two components), leading to a total of six pointwise deformation modes of the beam. 
The implementation of the Simo-Reissner beam finite element utilized in this work is detailed in~\cite{meier_geometrically_2019}. 
This reference also provides a comprehensive overview of the geometrically exact beam theory in general.

The radial support of a stent after the implantation into a diseased coronary artery section relies on the plastic deformation of the stent structure. 
The stent material is assumed as Co-20Cr-15W-10Ni, as this alloy is often used for coronary stents. 
To model the elasto-plastic behavior of the material, we propose a rate-independent plasticity model integrated into the geometrically exact beam theory used to model the stents. 
The constitutive relation between deformation measures and stress resultants, accounting for elasto-plastic material behavior, is detailed in~\ref{sec:app_elasto-plastic}. 
In current stent designs, the global radial extension of the stent is governed by local bending deformation of the struts. 
Therefore, we only model an elasto-plastic behavior of the bending moment to bending strain relationship within the beam theory following a linear isotropic hardening law. 
It is important to note that even though only the bending deformation is modeled as being elasto-plastic, the struts can still exhibit (elastic) torsion, shear, and axial extension deformations (see~\ref{sec:app_elasto-plastic} for details). 
The Young's modulus is \revs{$E = \unit[380]{GPa}$}, and the elasto-plastic tangent modulus $E^{ep} = \unit[64]{GPa}$~\cite{noauthor_materials_2009}. 
The initial yield moment $M_y^0$ was assumed to be reached once the normal stress on the outer fibers of the beam equaled the initial yield stress $\sigma_y$, thus \revs{$M_y^0 = \sigma_y \pi R^3 / 4  = \unit[8 {\times} 10^{-5}]{Nm}$}, where $R = \unit[0.06]{mm}$ is the beam cross-section radius. 
\reva{We directly start our simulation with the stent in a crimped configuration on the balloon catheter, i.e., we assume a stress-free initial configuration.  
To account for the kinematic hardening from the stent crimping, we modified the Young's modulus and the initial yield moment from the material parameters in~\cite{noauthor_materials_2009} such that the stent expansion and elastic recoil match the manufacturer's data for this stent type~\cite{biomatrix_stent} and the mean elastic recoil based on clinical studies~\cite{aziz_stent_2006}.}

\subsection{Simplified balloon model}
\label{sec:balloonmodel}

In this work, we model the balloon as a simplified, elastically deforming thin-walled tube and neglect the folding and pleating typical for coronary balloon catheters. 
To nevertheless achieve a realistic inflation behavior, we describe the balloon material as an artificially orthotropic material with an exponential material model in the two preferred anisotropy directions governing the inflation behavior, as proposed in~\cite{kiousis_NumericalModelStudy_2007}.
The balloon is spatially discretized with one layer of 3D hexahedral finite elements with linear shape functions. 
It is well known that this element type can be prone to shear locking. 
However, over almost the entire balloon length its overall behavior is primarily governed by membrane deformations.
Large shear states only occur localized at the balloon ends.
Since the balloon serves as an idealized structure to control the stent expansion, we can accept some artificial stiffening at the balloon ends and choose the material parameters accordingly.
The material is a hyperelastic Neo-Hooke base material with exponentially stiffening anisotropic contributions as described in~\ref{sec:app_strainenergies}.
We choose the balloon material parameters to obtain a realistic balloon inflation behavior, i.e., the balloon diameter at the respective inflation pressure, as we will demonstrate in Section~\ref{sec:freestentexpansion}.
The Neo-Hookean base material possesses a Young's modulus of $\unit[17.0]{MPa}$ and a Poisson's ratio of $0$.
We define the artificial fiber parameters $k_1 = 1000$, $k_2 = 0.01$ for the longitudinal direction and $k_1 = 1.5 {\times} 10^{-7}$, $k_2 = 0.35$ for the circumferential direction.
We add a small material dissipation term to numerically stabilize buckling phenomena as described in Section~\ref{sec:case2results}.
The mathematical formulation of the boundary value problem is similar to Equation~\eqref{eq:artery_bvp}.
In this case, the inner pressure on the Neumann boundary is the balloon inflation pressure $p_{in}$.
The balloon ends are fixed using spring boundary conditions.
We found that this set of boundary conditions is the best approximation for the connection of the balloon ends with the coronary catheter.
Additionally, it reduces strong element deformations at the boundary elements compared to fixed Dirichlet boundary conditions.

This simplified balloon model helps to significantly reduce the computational cost of the total simulation while maintaining realistic results for the stent expansion. 
It also allows for the replication of typical effects seen in the folded balloon expansion.
First, with increasing inner pressure, the material unfolds initially at its proximal and distal ends before progressing towards the middle section, the so-called dogboning effect~\cite{rahinj_numerical_2021}.
A uniform diameter across the full balloon length characterizes the resulting configuration at the nominal inflation pressure. 
Additionally, when placing the balloon catheter in a curved artery section, the fully inflated balloon shows a reduced curvature. 
This post-stenting straightening effect was also reported in \textit{in vivo} studies~\cite{wentzel_CoronaryStentImplantation_2000a, liao_ThreeDimensionalAnalysisVivo_2004} and simulations with fully resolved, folded balloon models~\cite{chiastra_computational_2016, mortier_novel_2010, morlacchi_patient-specific_2013}.
Balloon kinking often happens during inflation in very curved sections, as shown for aortic stent grafts in~\cite{mandigers_utilizing_2023} and can also be modeled with this approach. 
The artificially orthotropic balloon material formulation used in this study captures these typical effects, unlike approaches where displacement is prescribed purely radially.

\subsection{Initial configuration of the stenting device}
\label{sec:prebending}

Coronary stenosis often affects long artery sections, requiring stent lengths of up to $\unit[60]{mm}$. 
Due to their high curvature, especially in the right coronary artery, placing the stent into a curved artery section is inevitable. 
The relatively low bending stiffness of the uninflated balloon-stent system allows for the bending of the system to place the stent in the correct position. 
We neglect the procedure of the stenting device positioning to the target site and directly start the simulation from a predefined, bent configuration inside the vessel. 
For the final mechanical load on the artery, the exact initial configuration is not critical.
Hence, \revb{we propose a method to estimate an initial configuration for the stenting device that enables the application to curved artery sections without the necessity to simulate the catheter insertion process.}

\begin{figure}[h]
    \centering
    \includegraphics[width=\textwidth]{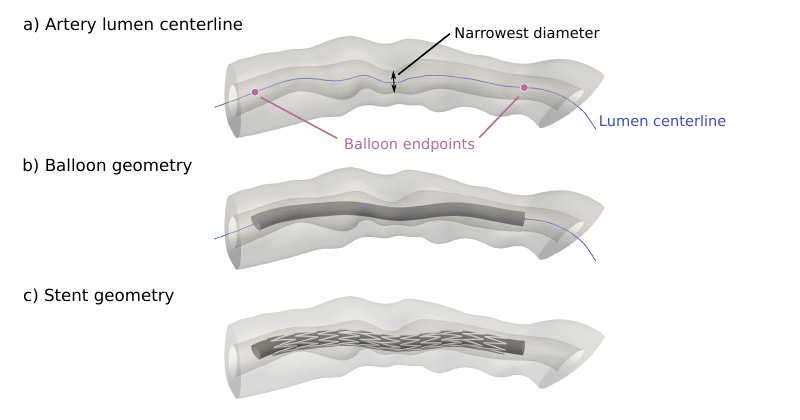}
    \caption{Patient-specific artery geometry with lumen centerline, initial balloon and stent geometry}
    \label{fig:ballooncenterline}
    \end{figure}

In the first step, the balloon geometry in the stenosed area before the inflation is determined.  
Figure~\ref{fig:ballooncenterline}a) shows an example of the artery geometry with its lumen centerline and the location of the narrowest diameter, which are the basis for the balloon geometry and position.
We assume the stent is placed centrally at the stenosis, i.e., the stent midpoint is at the narrowest diameter site. 
The endpoints are determined with the targeted balloon length and are assumed to lie on the lumen centerline.

We use ScyPy's univariate spline fit~\cite{2020SciPy-NMeth, dierckx1995curve} to interpolate the lumen centerline points between the balloon endpoints to obtain the balloon centerline. 
The number of knots to interpolate the data is increased until the smoothing condition 
\begin{equation}
    \Bigl| \sum\limits_{i=0}^{n} \left( (w_i \cdot (y_i-\mathrm{spl}(x_i)))^2 \right) - \zeta \Bigr| < \epsilon_\zeta \cdot \zeta 
\end{equation}
for a cubic spline $y = \mathrm{spl}(x)$ with the smoothing factor $\zeta$ is satisfied under a tolerance of $\epsilon_\zeta = 0.001$.
Here, $n$ is the number of centerline points $(x, y)$. 
We use a high smoothing factor of $\zeta = 4$ for the interpolation in all three coordinate directions to ensure a relatively straight balloon geometry.
An array of adapted weights $w$ for the spline fitting ensures no overlap of the balloon geometry with the endothelial surface emerges.
We create the balloon geometry as a hollow cylinder along this centerline; the outer diameter is set slightly smaller than the narrowest artery diameter. 
Figure~\ref{fig:ballooncenterline}b) shows the final balloon geometry for this example. 

The creation of the initial stent geometry is done in two steps.
First, the stent is modeled as a straight device, which in the second step is warped onto the balloon geometry, see Figure~\ref{fig:ballooncenterline}c).
This is done in such a way that the stent centerline is not stretched during this warping procedure.
The rotational aspect of the transformation is chosen to minimize the twist along the deformed stent centerline.
This mapping depicts a preparatory step of the simulation; we then use the deformed but stress-free configuration as the initial configuration for the simulations.

\subsection{Contact mechanics}
\label{sec:beamcontact}

In the considered simulations, two types of contact interactions occur: surface-to-surface contact and beam-to-solid contact.
The surface-to-surface interaction appears between the balloon and the inner artery wall.
For this interaction, we employ a mortar finite element framework for contact mechanics, as introduced in~\cite{popp_FiniteDeformationMortar_2009}. 
As constraint enforcement strategy for the frictionless contact, a Lagrange multiplier approach with node-wise penalty regularization is performed.
Mortar methods for surface-to-surface contact are well established and will not be described in more detail here.
The interested reader is referred to the aforementioned reference and~\cite{popp_FiniteDeformationMortar_2009, popp_dual_2010} for more information. 

The beam-to-solid contact occurs between the stent and the balloon, as well as between the stent and the artery wall.
From a geometrical point of view, this is a line-to-surface (1D-2D) contact problem, which requires to formulate a mixed-dimensional contact interaction between the beams and the solid.
To model this mixed-dimensional interaction between 1D Cosserat beams and 2D solid surfaces, we use a novel approach introduced by~\cite{steinbrecher_consistent_2022, steinbrecher_mixed-dimensional_nodate}.
In~\cite{steinbrecher_consistent_2022}, a coupling formulation, i.e., mesh-tying, between 1D beams and 2D surfaces is presented, which is then extended to unilateral frictionless contact in~\cite{steinbrecher_mixed-dimensional_nodate}. 
There, the contact constraints are formulated via the space-continuous penalty potential
\begin{equation}
    \Pi_{\epsilon} = \int_{\Omega^{\mathcal{B}}} \frac{\epsilon}{2} \langle g(s)\rangle^2 \, \mathrm{d}s
    \qquad \mathrm{with} \qquad  
    \langle x \rangle = \begin{cases}  x, & x \leq 0\\ 0, & x > 0\\ \end{cases} .
\end{equation}
Here, $\Omega^{\mathcal{B}}$ is the full beam domain, $\epsilon \in \mathbb{R}^{+}$ is the penalty parameter and $g$ is the gap function.
Contributions to the penalty potential solely come from negative values of the gap function, i.e., penetration, while positive values, i.e., separation, do not contribute to the penalty potential.
For this work, the contact geometry of the beam finite elements is assumed to be a tube along the beam centerline with a constant radius $R$.
This allows to define the gap function via a minimization problem
\begin{equation}
    g(s) = \min_{\xi, \eta} \norm{\boldsymbol{r}(s) - \boldsymbol{x}_s(\xi, \eta)} - R.
\end{equation}
Here, $\boldsymbol{r}$ is a point on the beam centerline, $\boldsymbol{x}_s$ is the corresponding closest point on the solid surface obtained with a closest point projection.
The surface parameter coordinates $\xi$ and $\eta$ describe the position of $\boldsymbol{x}_s$ on the solid surface.

The validity of modeling assumptions of mixed-dimensional interaction problems has been studied in detail for embedded fibers, cf.~\cite{steinbrecher_mortar-type_2020,steinbrecher_consistent_2022-1}, with the conclusion that the solid finite elements should not be smaller than the beam cross-section dimensions.
In the case of beam-to-solid contact, this restriction is even softer.
For all the examples presented in this work, the solid element dimensions exceed the beam cross-section dimensions, thus justifying the use of a mixed-dimensional contact scheme.


\section{Free stent expansion simulation}
\label{sec:freestentexpansion}

As a first numerical example, we show the free expansion of the stenting device as a benchmark for the balloon inflation and stent expansion. 
We present a simplified model consisting of a straight stenting device and a more complex example, which will be used for the generic and patient-specific angioplasty simulations in Section~\ref{sec:case1} and~\ref{sec:case2}, respectively.

\subsection{Straight stenting device}
\label{sec:straightstentingdevice}

\begin{figure}[h]
    \centering
    \includegraphics[width=\textwidth]{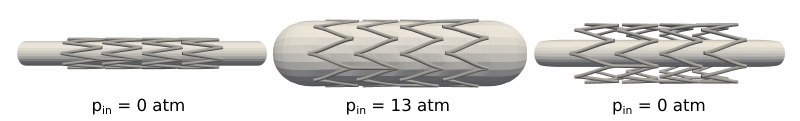}
    \caption{\revs{Free expansion simulation for the straight stenting device}}
    \label{fig:freeexpansion-generic}
    \end{figure}

We generate the stent and balloon models as described in Sections~\ref{sec:stentmodel} and~\ref{sec:balloonmodel} and sketched in Figure~\ref{fig:freeexpansion-generic}. 
For the first example, we use a straight stenting device with a stent length of $\unit[6]{mm}$.
The stent geometry comprises four rings. 
Each stent strut is discretized with three Simo-Reissner beam finite elements, totaling $210$ elements for the stent geometry.
The balloon's geometry is a straight cylinder with a length of \reva{$\unit[9.4]{mm}$}, a wall thickness of $\unit[0.04]{mm}$, and an outer diameter of $\unit[0.98]{mm}$.
It is discretized with \reva{$1{,}276$}  linear, hexahedral elements.
\reva{The balloon length in relation to the stent length was chosen such the tapers at both ends are sufficiently large to form the air pockets that cause the dogboning effect.}
In this configuration, the spring boundary stiffness at the balloon end surfaces is set to $\unit[0.1]{GPa}$.
The stent is fixed at four nodes as described in Section~\ref{sec:stentmodel} to only allow radial displacement. 

The inner pressure $p_{in}$ drives balloon inflation, which in turn expands the stent. 
We model the balloon and stent interaction as described in Section~\ref{sec:beamcontact} with a penalty parameter of $\epsilon = \unit[10]{N/mm^2}$.
The pressure is linearly ramped up with a maximum of $\unit[13]{atm}$ or $\unit[1.01]{MPa}$, which represents a typical value for commercially available coronary balloon catheters. 
After reaching the maximum pressure, the load linearly decreases such that the end-configuration is unloaded.
In total, $100$ quasi-static load steps are performed. 

\begin{figure}[h]
\centering
\includegraphics[width=\textwidth]{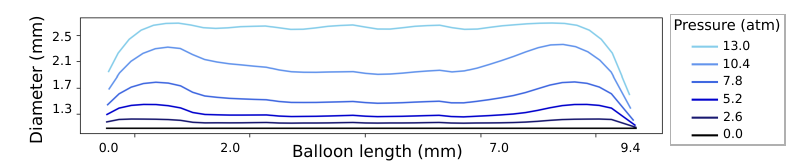}
\caption{\reva{Balloon diameter at different pressures for the generic free stent expansion simulation}}
\label{fig:diagramdogboning}
\end{figure}

Figure~\ref{fig:diagramdogboning} illustrates the balloon shape and diameter for different pressures during the free stent expansion simulation. 
The expansion shows the typical pattern for balloon unfolding, i.e., a rapid expansion with linear pressure increase, followed by a relatively constant diameter for pressures exceeding the balloon's nominal pressure. 
At the beginning of the inflation phase, the sections towards the balloon margins inflate at a higher rate than the middle section, causing a slight dogboning effect, which is also the case in the expansion of folded balloons.
\reva{In this example, the dogboning is most prominent during the inflation at \unit[10.4]{atm}, where the ratio between the distal and the central radius is $\unit[20.4]{\%}$. This is slightly lower than typical values for e.g. diamond-shaped stents~\cite{migliavacca_MechanicalBehaviorCoronary_2002}.}

Figure~\ref{fig:freeexpansion-generic} shows the free expansion simulations. 
The figure shows the balloon and stent initial configurations, the deformation at maximum pressure, and the final state of the balloon and stent after removing the pressure. 
Following an expansion from the initial diameter of \revs{$\unit[1.15]{mm}$}, the configuration at maximum pressure shows a rather uniform stent diameter over its length of approximately \revs{$\unit[2.59]{mm}$}. 
These numbers comply with typical values for commercially available stenting devices, such as the BioMatrix neoflex stenting system, which reaches a diameter of $\unit[2.71]{mm}$ at $\unit[13]{atm}$~\cite{biomatrix_stent}. 
The expansion behavior is also comparable with data obtained through experimental studies, cf.~\cite{geith_experimental_2020}.

The elasto-plastic properties of the stent material characterize its deformation during and after the balloon inflation. 
The stent deforms elastically up to a certain yield point and plastically beyond this point. 
Once the external force diminishes, i.e., the balloon deflates completely, the elastic part of the stent deformation is reverted and the plastic deformation remains. 
Hence, an elastic recoil reduces the post-procedure diameter from its maximum value.   
In this case, we observe a slight elastic recoil; the post-procedure diameter is \revs{$\unit[2.31]{mm}$}. 
We computed an absolute stent elastic recoil of \revs{$\unit[0.28]{mm}$}, 
i.e., \revs{$\unit[10.8]{\%}$}.  
Clinical studies suggest a mean elastic recoil of $\unit[10.02]{\%}$ with a standard deviation of $\unit[5.87]{\%}$, depending on the initial configurations and stenting devices used~\cite{aziz_stent_2006}.

\subsection{Curved stenting device}
\label{sec:patspec_stentingdevice}

In the next step, the free expansion simulation is applied to a more realistic stenting device setup, i.e., the geometry follows a curved centerline adapted to the patient-specific artery geometry, cf. Section~\ref{sec:prebending}. 
The stent length is $\unit[16.5]{mm}$.
In this example, the stent geometry comprises 11 rings and consists of $588$ Simo-Reissner beam finite elements.
The balloon has a similar wall thickness and diameter as in Section~\ref{sec:straightstentingdevice}; the length is $\unit[18]{mm}$.
The total element number for the full-length balloon is $2{,}765$.
In this example, we choose a spring stiffness of $\unit[1]{GPa}$ for the cut-off surfaces.

\begin{figure}[h]
\centering
\includegraphics[width=\textwidth]{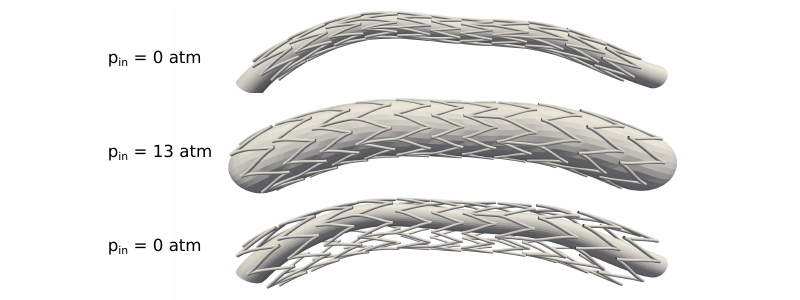}
\caption{\revs{Free expansion simulation for the stenting device}}
\label{fig:freeexpansion-patspec}
\end{figure}

Figure~\ref{fig:freeexpansion-patspec} shows the free expansion simulation.
Since the pressure is applied evenly to the inner balloon surface, the resulting stent diameters are similar to the straight setup. 
In the curved stenting device, we furthermore observe the post-stenting straightening effect from the balloon inflation. 
As visible in Figure~\ref{fig:freeexpansion-patspec}, the stenting device centerline straightens when the balloon inflates and also remains in a straighter configuration once the balloon pressure is removed.

\section{Coronary angioplasty simulation}
\label{sec:numex}

In the following, we apply the proposed approach to coronary angioplasty simulations. 
Case 1 shows a simplified example, with a generic cylinder geometry representing the artery. 
In Case 2, we use a patient-specific setup with a strong curvature to illustrate the applicability to complex models.

\subsection{Case 1: Generic artery model}
\label{sec:case1}

In this example, we use a simplified cylindrical artery to show the interaction between the balloon, stent, and artery. 
Although the artery is not stenosed, it expands through the stent placement.

\subsubsection{Simulation setup}
\label{sec:simulationsetup}

\begin{figure}[h]
    \centering
    \includegraphics[width=\textwidth]{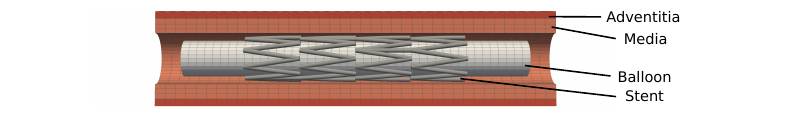}   
    \caption{\revs{Setup of the generic artery angioplasty simulation}}
    \label{fig:case1setup}
\end{figure}

The artery model is a generic hollow cylinder (generated using Coreform Cubit, 2018) with a two-layered nonlinear, hyperelastic, anisotropic material model as described in Section~\ref{sec:arterymodel}. 
We assume the intima is mechanically irrelevant for this preliminary example, and hence, we only describe the media and adventitia. 
The probe length in this example is \revs{$\unit[11]{mm}$}.
The outer diameter of $\unit[2.8]{mm}$ and the total wall thickness of $\unit[0.6]{mm}$ are within the range of a normal left coronary artery section~\cite{holzapfel_determination_2005}. 
We use a structured, hexahedral mesh with linear basis functions for the spatial discretization of the simplified artery model. 
The stenting device, including balloon and stent, is modeled as described in Section~\ref{sec:straightstentingdevice}.
The full setup is shown in Figure~\ref{fig:case1setup}. 
It holds \revs{$33{,}960$} \revss{solid} degrees of freedom \revss{(DOFs; $6{,}330$ for the balloon and $27{,}000$ for the artery) and $3{,}150$ beam DOFs.}

Similarly to Section~\ref{sec:freestentexpansion}, the balloon inflation governs the stent expansion. 
The stent interaction with the balloon catheter and the artery is modeled using a mixed-dimensional beam-to-solid contact formulation, cf. Section~\ref{sec:beamcontact} with a penalty of $\epsilon = \unit[10]{N/mm^2}$~\cite{steinbrecher_mortar-type_2020,steinbrecher_consistent_2022,steinbrecher_consistent_2022-1,steinbrecher_mixed-dimensional_nodate}.
Contact mechanics between the balloon and artery is described using a mortar finite element framework for 3D surfaces~\cite{popp_FiniteDeformationMortar_2009, popp_dual_2010}. 
The penalty parameter for the surface-to-surface interaction is set to $\epsilon^\mathcal{S} = \unit[5]{N/mm^2}$, allowing for minimal penetration while maintaining computational robustness and speed.
Since it is not the primary interest of this manuscript, we neglect the exact artery prestressing and resort to the application of an appropriate typical diastolic pressure of $p_{d} = \unit[85]{mmHg}$.
At the cut-off surfaces, we use omnidirectional springs with a stiffness of $\unit[1]{GPa}$ as boundary conditions.
\revb{We found that this value represents the tissue of the adjacent tissue well for this example, as the cut-off surfaces do not affect the stresses measured in the stented area.}

\subsubsection{Results}

\begin{figure}[h]
\centering
\includegraphics[width=\textwidth]{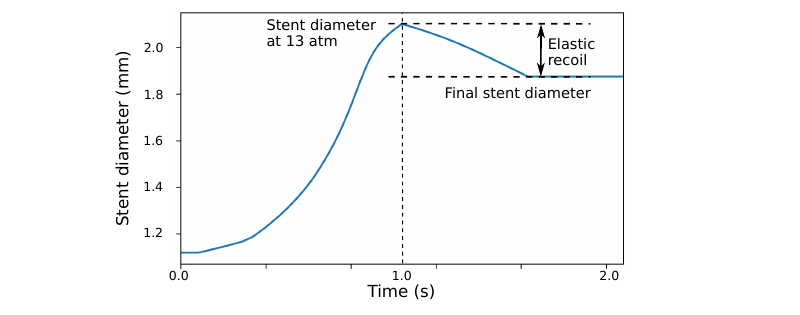}   
\caption{\revs{Stent diameter over balloon inflation and deflation}}
\label{fig:stent_diameter}
\end{figure}

Figure~\ref{fig:stent_diameter} plots the stent diameter over the balloon inflation and deflation. 
The horizontal axis is the (pseudo-)time of the simulation, corresponding to the linearly increased pressure to a maximum of $\unit[13]{atm}$ at time $\unit[1.0]{s}$, decreasing linearly again to $\unit[0]{atm}$ at time $\unit[2.0]{s}$.
The diameter is computed from the mean radial displacement of all stent nodes. 
The stent diameter increases rapidly during the early inflation phase, followed by a reduced diameter increase towards the maximum pressure. 
At the maximum balloon pressure of $\unit[13]{atm}$, the stent diameter is \reva{$\unit[2.12]{mm}$}. 
After the elastic recoil \reva{of $\unit[10.8]{\%}$}, the final stent diameter is \reva{$\unit[1.89]{mm}$}. 

\begin{figure}[h]
\centering
\includegraphics[width=\textwidth]{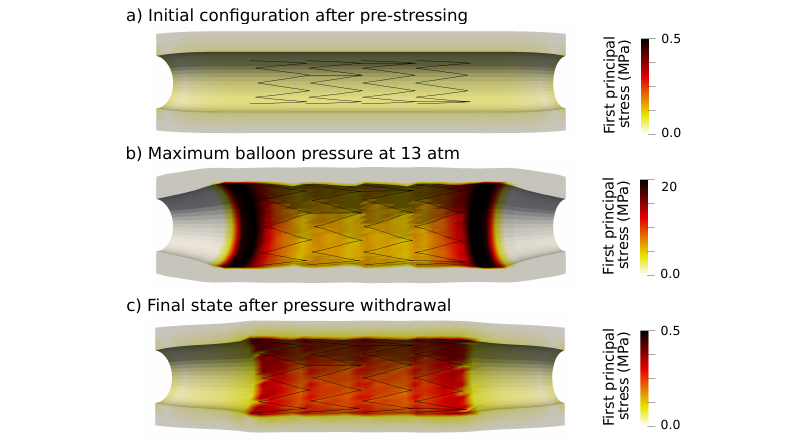}   
\caption{\revs{Stent deployment simulation in a generic artery model; for a better overview, the balloon is not shown, and the stent is outlined}}
\label{fig:genericartery}
\end{figure}

Figure~\ref{fig:genericartery} shows the generic coronary artery setup in its initial configuration, at full expansion with $\unit[13]{atm}$ balloon pressure, and after the pressure withdrawal. 
The colors indicate the first principal Cauchy stresses in the current configuration, a scalar quantity that is, in this case, mainly directed circumferentially. 
In the initial configuration, shown in Figure~\ref{fig:genericartery}a), no contact between the stent and the artery has been established yet.
After the prestressing with an inner artery pressure of $\unit[85]{mmHg}$, the maximum deformation is $\unit[0.14]{mm}$.
The respective first principal stress reaches \revs{$\unit[2.3]{kPa}$} and \revs{$\unit[0.17]{kPa}$} in the media and adventitia, respectively. 
Figure~\ref{fig:genericartery}b) shows the configuration of the maximum balloon pressure of $\unit[13]{atm}$. 
The highest principal stress values of \revs{$\unit[7.99]{MPa}$} are reached at the endothelial surface near the contact points with the stent struts. 
The first principal stresses in the adventitia reach \revs{$\unit[49.5]{kPa}$}. 
Over the length of the stented site, the sites proximal and distal to the stent experience slightly higher stresses than the rest of the endothelium. 
Figure~\ref{fig:genericartery}c) displays the stented generic artery after the withdrawal of the balloon pressure. 
The first principal stress is significantly lower and more evenly distributed than during maximum inflation. 
Maximum values for the media and adventitia are \revs{$\unit[187.5]{kPa}$}, and \revs{$\unit[13.7]{kPa}$}, respectively.

\subsection{Case 2: Patient-specific artery model}
\label{sec:case2}

In this section, the presented approach is applied to a patient-specific coronary artery setup to demonstrate the feasibility of working with arbitrarily shaped arteries as retrieved from medical imaging data.

\subsubsection{Patient-specific artery and simulation setup}

\begin{figure}[htbp]
    \centering
    \includegraphics[width=\textwidth]{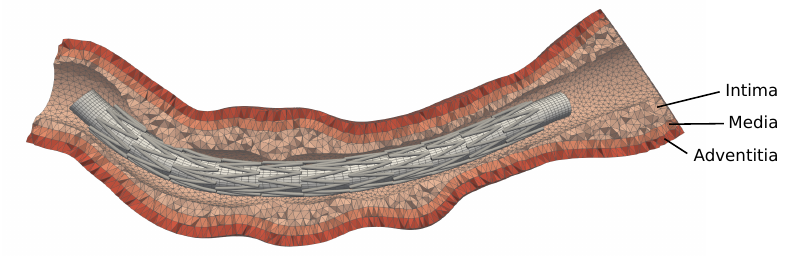}   
    \caption{\revb{Patient-specific artery with stenting device}}
    \label{fig:setuppatspec}
    \end{figure}

We use coronary computed tomography angiography (CCTA) image data obtained at the German Heart Center Munich from an exemplary patient diagnosed with coronary artery disease as a basis for the computational artery model.
The diseased area is localized at the proximal right coronary artery section. 
\revb{The lesion is identified as a non-calcified, mixed-tissue coronary plaque with approximately $\unit[20]{\%}$ luminal obstruction. Additionally, some positive remodeling, i.e., thickened intimal tissue leading to increased outer diameter, can be observed.}
The respective section, including the stenosis and the surrounding area, is segmented using the open-source software SimVascular~\cite{updegrove_simvascular_2017}. 
The length of the segmented artery section is approximately $\unit[21]{mm}$, and the inner diameter is between $1.5$ and $\unit[2.0]{mm}$.
The lumen centerline is computed and exported using the SimVascular reduced-order modeling tool~\cite{updegrove_simvascular_2017}.
We assume the outer layers of the artery wall, namely the adventitia and media, to have fixed thicknesses, which we calculate based on the ratio to the outer vessel diameter as found in literature~\cite{holzapfel_determination_2005}. 
The remaining tissue of the inner layer is defined as intimal tissue. 
The total wall thickness varies from $0.7$ to $\unit[1.8]{mm}$.
We model the artery material as described in Section~\ref{sec:arterymodel} and apply the patient-specific diastolic pressure of $p_{d} = \unit[85]{mmHg}$, since the CCTA image acquisition is typically gated in this phase.
We define the thickened intima as homogenous, neglecting the personalized composition of the diseased tissue and we assume its material properties resemble the media's. 
A harmonic lifting operation, as described by~\cite{nagler_PersonalizationCardiacFiber_2013}, provides a smooth fiber distribution throughout the complex domain of the patient-specific geometry using the lumen centerline as a construction basis for the local coordinate system. 
For this, we use the specific lumen centerline's tangent as the orientation of the longitudinal direction. 
We use the software package LNMmeshio~\cite{gebauer_lnmmeshio_2023} for the model preparation and Gmsh~\cite{geuzaine_gmsh_2009} for the mesh generation. 
A tetrahedral finite element discretization with quadratic basis functions is employed. 
In total, the artery mesh consists of $156{,}590$ elements \revb{and $238{,}123$ nodes; the mean mesh size is $\unit[0.25]{mm}$}. 
The artery layers are meshed separately and have a conforming mesh at their internal boundary surfaces.
\revb{Figure~\ref{fig:setuppatspec} shows the patient-specific artery components and discretization}.

The stent geometry is generated with a length of $\unit[18]{mm}$ and $11$ rings as described in Section~\ref{sec:stentmodel}.
We determine the initial configuration of the balloon and the stent as described in Section~\ref{sec:prebending}.
Similarly to the numerical example of the generic artery presented in Section~\ref{sec:case1}, the stent expansion is governed by the balloon inflation and the cut-off surfaces are supported by omnidirectional springs. 
The orthotropic material properties of the balloon allow a radial deformation and prevent the balloon from excessive bending under the inner pressure, cf. Section~\ref{sec:patspec_stentingdevice}. 
Figure~\ref{fig:patspec}a) shows the simulation setup.

\subsubsection{Linear solver and preconditioning}

The balloon, stent, and patient-specific coronary artery section contain  \revss{$16{,}800$, $8{,}811$, and $714{,}369$ DOFs.}
The simulation requires small load steps to ensure convergence through the highly nonlinear system behavior; we use $500$ steps in this example. 
Additionally, the material parameters reside in different orders of magnitude, and the system involves thin-walled structures and mixed-dimensional modeling; therefore, the resulting matrix is ill-conditioned. 
To accommodate the large problem size and high condition number, we employ an iterative Krylov-based solver with specialized, problem-oriented preconditioning for a fast and efficient solution. 
We use the GMRES implementation from the Trilinos software package Belos~\cite{sandia_national_laboratories_trilinosbelos_nodate}.
We use an Incomplete LU (ILU(p)) preconditioner with a Krylov subspace of size $200$ and a fill level $p = 4$ to improve the convergence. 
The implementation in 4C~\cite{wall_baci_2018} allows for MPI parallelism. 
The computation of the current example is distributed on 20 MPI processes.

\subsubsection{Results}
\label{sec:case2results}

\begin{figure}[htbp]
\centering
\includegraphics[width=\textwidth]{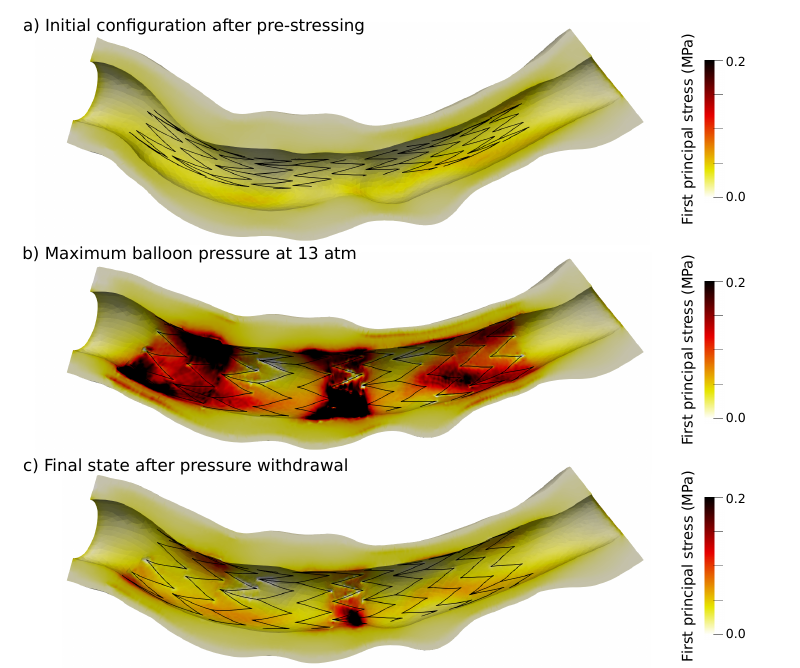}   
\caption{\revs{Stent deployment simulation in a patient-specific model. For better visibility, the balloon is not shown.}}
\label{fig:patspec}
\end{figure}

Figures~\ref{fig:patspec}b) and c) show the setup with the fully inflated balloon and the end-state after the pressure withdrawal, respectively. 
The modeled stent implantation successfully reopened the narrowed artery section. 
The colors indicate the first principal Cauchy stress values.

As we have observed in Case 1, the first principal stress in the artery wall reaches its highest value at maximum balloon inflation. 
The endothelial tissue in direct contact with the stent experiences high principal stresses of up to \revs{$\unit[757.0]{kPa}$}, which is a \revs{$15$}-times increase from the initial prestressed value of \revs{$\unit[48.1]{kPa}$}. 
In comparison, the value decreased to \revs{$\unit[191.4]{kPa}$} after PCI. 
At maximum balloon inflation, we see stress values of up to \revs{$\unit[78.3]{kPa}$} and \revs{$\unit[135.9]{kPa}$} in the media and adventitia, respectively. 
Those values reduce to \revs{$\unit[48.9]{kPa}$} and \revs{$\unit[49.5]{kPa}$} after the intervention. 
Hence, the principal stresses after PCI are elevated from the initial configuration with a factor of \revs{$4.0$}, \revs{$1.4$}, and \revs{$3.9$} in the intima, media, and adventitia, respectively, potentially leading to long-term growth and remodeling effects.

Due to the complex shape of the image-based model, the stress distribution is irregular throughout the artery. 
In severely stenosed areas, the intima and media principal stress values during and after PCI are higher compared to less stenosed parts. 
Comparably high media stress values are also reached in the proximal and distal sites of the stenosis. 
\revb{The first principal stresses are oriented in the circumferential direction both during maximum inflation and after the balloon withdrawal, i.e., circumferential stresses exceed the radial contact forces from the stent struts.
In~\ref{sec:app_results}, the directions of the principal stress vectors are visualized.}
The locations of high stresses in this numerical example might indicate areas of high risk for ISR, and they correlate with clinical observations regarding frequent ISR locations. 
Overall, the stresses on and near the endothelial surface in contact with the stent exceed those in the remaining tissue. 
In contrast, the media tissue towards the adventitia experiences lower stresses than the adventitia itself.  
The adventitia serves as a protective layer for the artery against overstretch and rupture as it becomes excessively stiff when experiencing high loads~\cite{holzapfel_determination_2005}.

The final stent configuration is in place of the stenosed section and holds an opened position after plastic deformation. 
The stent implantation slightly straightens the curved artery section; some curvature is retained after the pressure withdrawal. 
We observe an instability mode in the balloon model shortly before the diameter reaches its maximum value, which results from compression forces on the inner side of strongly curved balloon sections. 
Load-carrying thin-walled structures are especially sensitive to buckling when bending is involved~\cite{ramm2002shells}.
As can also be observed in fully resolved balloon inflation simulations~\cite{mortier_novel_2010,mandigers_utilizing_2023}, the balloon buckles inwards before it reaches a stable state again at higher pressures. 
In less curved artery sections, we did not observe this behavior.

\subsection{Numerical performance}

We performed the simulations of the free expansion for the straight and the curved geometry on a Linux workstation using two or four parallel processors, respectively. 
Each processor features 62 GB of RAM 
and 16 11th Gen Intel Cores i9-11900 with $\unit[2.50]{GHz}$ operating frequency. 
We used $100$ load steps for the free expansion simulations, which took \revs{$\unit[14.7]{min}$}, and \revs{$\unit[31.6]{min}$} for the straight and the curved stenting device, respectively. 
We used four processors to simulate the generic artery angioplasty in Case 1. 
The simulation took \revs{$\unit[6.8]{h}$} for $500$ load steps. 
For the patient-specific example in Case 2, we used 20 cores on an Intel Xeon Gold 6230 ``Cascadelake'' processor with 96GB of RAM.
For $500$ load steps, this example took \revs{$\unit[23.4]{h}$}.

\section{Discussion}
\label{sec:discussion}

Mechanical factors are often overlooked as risk factors for in-stent restenosis after percutaneous coronary interventions. 
This manuscript proposes a novel \revbb{mixed-dimensional} approach for the patient-specific assessment of stenotic lesions during and after PCI. 
\revbb{While this work focuses on the application for PCI simulations, the methodology is versatile and can easily adapted to various problems in biomechanics or engineering that involve slender geometries, e.g., endovascular techniques for intracranial aneurysms~\cite{frank2024numerical,bisighini_MachineLearningReduced_2023}, or transcatheter aortic valve implantation procedures~\cite{grossi_ValidationEvidenceExperimental_2024}.}

\subsection{Simulation framework}

In this work, we combined methods for mixed-dimensional contact interaction, an elasto-plastic formulation for geometrically exact Simo-Reissner beams, and specifically designed materials for simplified balloon modeling with real-world patient-specific artery models. 
The proposed stenting device model is simple and easy to adapt for various complex geometries, as is the case for real-world applications. 
It fulfills the requirements for a realistic expansion behavior independent from its initial configuration. 
As described in Section~\ref{sec:freestentexpansion}, it expands evenly for realistic pressure values, showing typical effects like dogboning and post-stenting straightening. 
The maximum balloon diameter can be easily adapted to the lesion-specific case by altering the balloon material parameters or initial configuration.
Through automated geometry generation, the stent geometry can also be adjusted for different lesions or to compare the outcome using different stents.

\subsection{\revbb{Efficiency of the mixed-dimensional approach}}

The reduced-dimensional 1D stent model requires significantly less computational effort than fully resolved 3D structures. 
Nonetheless, it provides exact solutions during the interaction with the artery model through sophisticated methods for mixed-dimensional contact mechanics.  
\revbbrem{This approach focuses on the effect of the stent deployment on the arterial tissue; hence, the majority of DOFs belong to the artery model.}

\revbb{To get an idea of the computational effort saved by the mixed-dimensional approach, we present some conservative analytical considerations on the computational efficiency compared to a fully resolved stent model.
First, we consider the number of DOFs required for a fully resolved 3D stent model.
The stent struts have a diameter of $\unit[0.12]{mm}$ and a length of $\unit[1.52]{mm}$.
For simplicity, we assume a structured rectangular discretization of the strut cross-section with $5\times 5$ elements.
If we allow a maximum solid element length-to-width ratio of $2$, a fully resolved stent discretization would need $32$ element rows per stent strut.
This discretization aligns with ones reported in the literature, e.g.~\cite{morlacchi_patient-specific_2013,conway_computational_2012}.
However, we still consider this a conservative estimate, as this discretization might not yield a converged solution, especially considering the elasto-plastic material behavior of the stent.
Neglecting that an even finer discretization would be needed for the joints, this would yield approximately $155{,}000$ elements and $700{,}000$ DOFs for the whole stent.}

\revbb{The presented patient-specific model consists of $731{,}169$ solid DOFs ($714{,}369$ for the artery and $16{,}800$ for the balloon) and $8{,}811$ beam DOFs.
Thus, a fully resolved stent would nearly double the total number of DOFs for the system.
Even assuming perfect scaling behavior of the linear solver time with the number of DOFs in the presented numerical simulation framework, this would result in an increase in linear solver time by $\unit[100]{\%}$.}

\revbb{Another important aspect to consider is the numerical effort associated with evaluating the contact terms.
We assume that the contact evaluation time scales approximately linearly with the number of contact evaluation points, cf.~\cite{farahSegmentbasedVsElementbased2015}. 
The numerical effort associated with a single contact evaluation point is approximately equal for the beam-to-solid contact formulation and a surface-to-surface contact formulation, as both require a closest point projection and a linearization of the gap function.
For this comparison, we consider element-based integration schemes, as they allow for a more general estimate of the contact points, as opposed to segment-based integration schemes, which are highly dependent on the current configuration.
For the mixed-dimensional contact formulation, we assume six integration points per beam element.
Given that we have contact between the beam and the artery, as well as the beam and the balloon, this results in a total of $7{,}000$ contact evaluation points for the mixed-dimensional contact formulation.
For the fully resolved stent discretization, one side of the strut cross-section must be checked for contact with the artery and one for contact with the balloon.
Assuming $2\times 2$ contact evaluation points per surface element of the fully resolved stent model results in $248{,}000$ contact evaluation points.
This represents an increase of approximately $\unit[3{,}500]{\%}$, which is expected as the contact changes from a line-to-surface to a surface-to-surface formulation.
Considering that, in our simulations, the mixed-dimensional contact evaluation time was roughly $30\%$ of the linear solve time, this increase in contact evaluation points would significantly impact overall performance.}

\revbb{Even without regarding nonlinear solver aspects and memory bound performance aspects, these considerations show that a fully resolved 3D model, compared to the proposed mixed-dimensional model, would result in substantially higher computational costs (a multiplication of compute time or an increase of at least one order of magnitude) that are difficult or even impossible to meet.}

\subsection{Implications for in-stent restenosis}

To demonstrate the clinical applicability, we presented the assessment of mechanical properties during and after PCI with stenting for an exemplary patient-specific geometry in Section~\ref{sec:case2}.
We assessed the first principal Cauchy stress distribution and peak values during maximum balloon inflation and after balloon retrieval. 
We review the first principal stresses as the primary relevant mechanical value for vascular growth and remodeling~\cite{humphrey_MechanicsMechanobiologyModeling_2012, holzapfel_ComputationalApproachesAnalyzing_2014}. 
We found that principal stresses are significantly elevated during the maximum balloon inflation and mildly elevated after the balloon retrieval.
Long-term growth and remodeling are promoted by a persistent interference of the tissue's mechanobiological equilibrium~\cite{cyron_HomogenizedConstrainedMixture_2016}, which could be observed in multiple areas after the stent implantation. 
Acute tissue overstretch might additionally cause collagen damage, which can lead to a cascade of inflammatory processes promoting maladaptive growth~\cite{gierig_PostangioplastyRemodelingCoronary_2024}. 
Hence, these areas could be considered at a higher risk of ISR. 
This finding agrees with recent machine learning studies highlighting the balloon-to-vessel ratio as a risk factor relevant for ISR~\cite{guldener_machine_2023}. 
For this exemplary coronary artery model, we observed high stresses at the stent margins and very stenosed sites. 
Literature shows these locations to be frequently susceptible to ISR in drug-eluting stents~\cite{buccheri_understanding_2016, shafiabadihassani_InStentRestenosisOverview_2024}.
We aim to use such insights in further studies to determine potential sites at risk of damage and ISR.

\subsection{Strengths and limitations}

While many computational studies on PCI only include generic or simplified geometries, focus on the stent mechanics, or deploy computationally expensive stenting device models, we propose an approach suitable for complex, patient-specific geometries at relatively low modeling and computational effort.

However, there are limitations to the current approach. 
The presented patient-specific geometry is a preliminary model focusing on correctly representing the geometric shape.
Empirical literature data was used for the layer thicknesses, collagen architecture, material parameters, and boundary conditions. 
We did not categorize the intimal layer as healthy intima or plaque but used the same material properties as the media. 
We plan to analyze different plaque compositions and morphologies in a follow-up study. 
Furthermore, employing an appropriate prestressing algorithm will make the numerical results for the patient-specific cases more accurate.

The stent geometry was simplified by directly connecting the nodes within a ring. 
\revb{The local stent indentation into the endothelial surface, as is was e.g. shown in~\cite{geithQuantifyingStentinducedDamage2020}, was not explicitly modeled in our approach since we focused on the global system behavior rather than the fully resolved strut-level contact.}
Additionally, we neglected the stent axial and torsional plasticity since they play minor roles in the expansion. 
\revb{The initial configuration of the stenting device was estimated from the lumen centerline; to obtain the exact starting configuration, the catheter insertion process would have to be simulated separately.}
\reva{Furthermore, simplifying the overall stenting process simulation, i.e., adapting the models and material parameters of balloon and stent such that their behavior is realistic, rather than simulating the balloon folding, pleating, and stent crimping, introduces a small error in the expansion behavior and final configurations~\cite{geith_importance_2019}.}

\revbb{The current implementation of the presented approach is not optimized for efficiency yet and huge gains can still be achieved only via optimizing the implementation. In addition,}
the patient-specific simulations required a relatively small load step size to achieve robust convergence, leading to long overall simulation times. 
The simulation speed could be further increased by employing an adaptive step size control.
\revbb{Additionally, the computation time per load step could be decreased by using a specialized block factorization preconditioner, which is part of ongoing research~\cite{firmbach2024approximate}.}

\subsection{Clinical relevance and future research}

In the future, this approach can serve as a basis for further studies to quantify the long-term effects of mechanical changes from PCI. 
\revb{The important contribution of this manuscript is an efficient tool that allows to include one additional, and in our opinion crucial, aspect, namely the local stresses, as those have been ignored in large patient studies so far.}
For instance, identifying high-stress locations during and after PCI may provide further insight into the relation between specific lesions and their clinical outcomes by adding crucial physics information to the limited available data.
\revb{Specifically, comparing the influence of different plaque tissue, such as calcified, fibro-fatty, or mixed plaque, as well as different features, such as positive remodeling, eccentricity, or high tortuosity can provide new insights for the risk assessment of ISR.}
By including the characteristics of preprocedural coronary atherosclerotic lesions derived from CCTA, we aim to identify groups posing a specifically high ISR risk, in addition to the already known risk factors for ISR. 
Through the computational simulation of different stenosed arteries, we aim to find connections and explain and extend correlations identified by statistical and machine learning studies. 
\revb{In consecutive work towards the application to clinical practice, additional pre- and post-interventional imaging data, such as angiography data, can serve as a basis for further model verification.}

\newpage

\appendix

\section{Solid mechanics}
\label{sec:app_solidmechanics}

We model the artery and the balloon as 3D Boltzmann continua with a classical continuum solid mechanics approach, cf.~\cite{holzapfel_nonlinear_2002}. 
Mechanical quantities of interest are the deformation, strain, and stress within the tissue in the current configuration. 
The right Cauchy-Green tensor $\boldsymbol{C} = \boldsymbol{F}^T \boldsymbol{F}$ is computed from the deformation gradient $\boldsymbol{F} \in \mathbb{R}^{3 \times 3}$, which is defined as the partial derivative of the current configuration $\boldsymbol{x} \in \mathbb{R}^{3}$ with respect to the spatial configuration $\boldsymbol{X} \in \mathbb{R}^{3}$ of a material part on the solid body, i.e., $\boldsymbol{F} = \frac{\partial \boldsymbol{x}}{\partial \boldsymbol{X}}$.
The finite displacement field is further defined as $\boldsymbol{u} = \boldsymbol{x} - \boldsymbol{X}$.
Furthermore, the material Green-Lagrange strain tensor  $\boldsymbol{E}$ is obtained with 
\begin{equation}
    \boldsymbol{E} = \frac{1}{2} (\boldsymbol{C} - \boldsymbol{I}) ,
\end{equation}
where $\boldsymbol{I} \in \mathbb{R}^{3 \times 3}$ is the 3D second-order identity tensor.
The second Piola-Kirchoff stress tensor $\boldsymbol{S}^{\revss{\infty}} \in \mathbb{R}^{3 \times 3}$ relates to the Green-Lagrange strain via the strain energy function $\Psi^{\revss{\infty}}$, i.e.,  
\begin{equation}
    \boldsymbol{S}^{\revss{\infty}} = \frac{\partial \Psi^{\revss{\infty}}(\boldsymbol{E})}{\partial \boldsymbol{E}}   .
\end{equation}

\section{Artery constitutive equations}
\label{sec:app_strainenergies}

For the anisotropic material formulation used in the soft tissue of the artery, the collagen contribution to the material behavior resembles directional fiber families embedded in a matrix material~\cite{cowin_new_2004}. 
The isotropic matrix material is decoupled from the embedded collagen fiber families using an additive split for the \revss{hyperelastic} strain-energy function
\begin{equation}
    \Psi^{\revss{\infty}} (I_1, I_4, I_6) = \Psi_{iso}(I_1) + \Psi_{aniso}(I_4, I_6)   .
\end{equation}
Here, $I_1 = \tr\boldsymbol{C}$ is an isotropic invariant of the right Cauchy-Green deformation tensor  $\boldsymbol{C}$~\cite{holzapfel_biomechanics_2017};
$I_4 = \boldsymbol{M}_f \cdot (\boldsymbol{C}\boldsymbol{M}_f)$, and $I_6 = \boldsymbol{M}_{f}' \cdot (\boldsymbol{C}\boldsymbol{M}_{f}')$ are invariants additionally depending on the two preferred directions of the anisotropy $\boldsymbol{M}_{f}$ and $\boldsymbol{M}_{f}'$, i.e., the two fiber families.
The neo-Hookean isotropic part reads
\begin{equation}
    \Psi_{iso}(I_1) = \frac{1}{2} G(I_1 - 3)   ,
\end{equation}
with $G$ as the shear modulus of the matrix material in its reference configuration, which can be computed from the Young's modulus $E$ and Poisson's ratio $\nu$ with 
\begin{equation}
    G = \frac{E}{2(1+\nu)}  \; .
\end{equation}
The exponential functions for the elastic fibers are 
\begin{equation}
    \Psi_{aniso}(I_4, I_6) = \frac{k_1}{2k_2} [\exp[k_2(I_4-1)^2]+\exp[k_2(I_6-1)^2]-2]   ,
\end{equation}
where $k_1$ and $k_2$ are material constants~\cite{cowin_new_2004}. 

\revaa{In the media and intima, an additional viscous term accounts for material damping.}
\revaa{A viscous contribution is added to the hyperelastic term based on the generalized Maxwell model, i.e., a dashpot in parallel with a spring~\cite{holzapfel_ViscoelasticModelFiberreinforced_2001,bel-brunon_NumericalIdentificationMethod_2014}}
\begin{equation}
    \revaa{\Psi = \Psi^{\infty} + \gamma^{vis}(\boldsymbol{C}, \boldsymbol{\Gamma}^{vis})  .}
\end{equation}
\revaa{Its dissipative potential $\gamma^{vis}$ depends on the right Cauchy-Green strain tensor $\boldsymbol{C}$ and the kinematic conjugate $\boldsymbol{\Gamma}^{vis}$ of the fictitious stress tensor $\boldsymbol{Q}$, expressed by}
\begin{equation}
    \revaa{\boldsymbol{Q} = -2 \frac{\partial \gamma^{vis}}{\partial \boldsymbol{\Gamma}^{vis}}  . }
\end{equation}
\revaa{The evolution of the stress tensor $\boldsymbol{Q}$ is governed by}
\begin{equation}
    \revaa{\dot{\boldsymbol{Q}} + \frac{1}{\tau} \boldsymbol{Q} = \beta \dot{\boldsymbol{S}}}
\end{equation}
\revaa{with $\tau = \unit[0.5]{s}$ as the relaxation time and $\beta = 0.3$ as the ratio between the viscoelastic and the hyperelastic branches.}
\revaa{The global stress tensor $\boldsymbol{S}$ is then given by}
\begin{equation}
    \revaa{\boldsymbol{S} = \boldsymbol{S}^{\infty} + \boldsymbol{Q}   .}
\end{equation}

\section{Elasto-plastic constitutive law for beam theory}
\label{sec:app_elasto-plastic}

For a purely elastic material behavior, the force and moment stress resultants of the beam are typically related to the work-conjugated deformation measures according to
\begin{align}
    \boldsymbol{F}^\mathcal{B}=\boldsymbol{C}_{\boldsymbol{F}} \cdot \boldsymbol{\Gamma}, \quad
    \boldsymbol{M}^\mathcal{B}=\boldsymbol{C}_{\boldsymbol{M}} \cdot \boldsymbol{\Omega}.
\end{align}
Here, the constitutive tensors $\boldsymbol{C}_{\boldsymbol{F}}$ and $\boldsymbol{C}_{\boldsymbol{M}}$ have the following diagonal structure
\begin{align}
    \boldsymbol{C}_{\boldsymbol{F}}= \boldsymbol{diag}[EA, G\bar{A}_2, G\bar{A}_3], \quad
    \boldsymbol{C}_{\boldsymbol{M}}= \boldsymbol{diag}[GI_T, EI_2, EI_3].
\end{align}
Here, $E$ and $G$ are the Young's modulus and the shear modulus, $A$, $\bar{A}_2$ and $\bar{A}_3$ are the cross-section and the two reduced cross-sections, $I_2$ and $I_3$ are the two principal moments of inertia, and $I_T$ is the torsional moment of inertia. In our work, this constitutive law has been extended to account for elasto-plasticity with isotropic hardening in the bending moments. For this purpose, only beams with circular cross-sections are considered, thus $GI_T=EI_2=EI_3=:EI$. Moreover, both the moment stress resultant $\boldsymbol{M}^\mathcal{B}=\boldsymbol{M}_t+\boldsymbol{M}_b$ and the rotational deformation measure $\boldsymbol{\Omega}=\boldsymbol{\Omega}_t+\boldsymbol{\Omega}_b$ are split into torsion and bending components according to
\begin{align}
    \boldsymbol{M}_t = (\boldsymbol{e}_1^T\boldsymbol{M})\boldsymbol{e}_1, \,\, 
    \boldsymbol{M}_b = \boldsymbol{M} -(\boldsymbol{e}_1^T\boldsymbol{M})\boldsymbol{e}_1 \,\,
    \boldsymbol{\Omega}_t = (\boldsymbol{e}_1^T\boldsymbol{\Omega})\boldsymbol{e}_1, \,\, 
    \boldsymbol{\Omega}_b = \boldsymbol{\Omega} -(\boldsymbol{e}_1^T\boldsymbol{\Omega})\boldsymbol{e}_1.
\end{align}
Here, $\boldsymbol{e}_1=[1,0,0]^T$ is the first base vector of the global Cartesian coordinate frame. While a purely elastic behavior is assumed for the torsion moment, i.e., $\boldsymbol{M}_t=GI_T\boldsymbol{\Omega}_t$, an elasto-plastic behavior is assumed for the bending moment as follows
\begin{align}
    &\text{Elasto-plastic split:} & & \boldsymbol{\Omega}_b= \boldsymbol{\Omega}_b^{el} + \boldsymbol{\Omega}_b^{pl}\,,\\[3pt]
    &\text{Elastic constitutive law:} & &\boldsymbol{M}_b= EI \boldsymbol{\Omega}_b^{el}\,,\\[3pt]
    &\text{Yield function:} & &\Phi= \|\boldsymbol{M}_b\| - M_y(\overline{\Omega}_b^{pl}) \leq 0\,,\\[3pt]
    &\text{Plastic flow rule:} & &\dot{\boldsymbol{\Omega}}_b^{pl}=\gamma \: \boldsymbol{e}\\
    &&&\text{with } \ \gamma \geq 0,\ \boldsymbol{e} = \frac{\boldsymbol{\Omega}_b^{el}}{\|\boldsymbol{\Omega}_b^{el}\|}\,,\\[3pt]
    &\text{Loading/unloading criterion:} & &\gamma\: \Phi=0\,.
\end{align}
Here, the plastic multiplier $\gamma$ represents the rate of plastic bending deformation and $\overline{\Omega}_b^{pl}$ is the accumulate plastic bending deformation according to
\begin{align}
\overline{\Omega}_b^{pl}(t)=\int_0^t \gamma (\tilde{t}) d\tilde{t}.
\end{align}
In the present work, isotropic hardening according to the following linear relationship between yield moment $M_y$ and accumulated plastic bending deformation $\overline{\Omega}_b^{pl}$ is assumed
\begin{align}
M_y(\overline{\Omega}_b^{pl})=M_y^0 + HI \overline{\Omega}_b^{pl}.
\end{align}
Here, $M_y^0$ denotes the initial yield moment before any plastic flow has occurred, and H is the hardening modulus. The latter can be related to the elasto-plastic tangent modulus $E^{ep}$ and Young's modulus $E$ according to
\begin{align}
    H = \frac{E^{ep}}{1 - E^{ep}/E}.
\end{align}
In the spatially and temporally discretized problem setting, the elasto-plastic material law is evaluated at every Gauss quadrature point of the spatial finite element discretization, and the plastic bending deformation increment per time step is determined using a classical return mapping algorithm.

\section{\revb{Further visualizations of the simulation results}}
\label{sec:app_results}

\revs{In the following, we provide additional result visualizations for the patient-specific example Case 2, suplementary to the results presented in Section~\ref{sec:case2results}.}

\subsection{\revb{Principal stress directions}}

\begin{figure}[htbp]
\centering
\includegraphics[width=\textwidth]{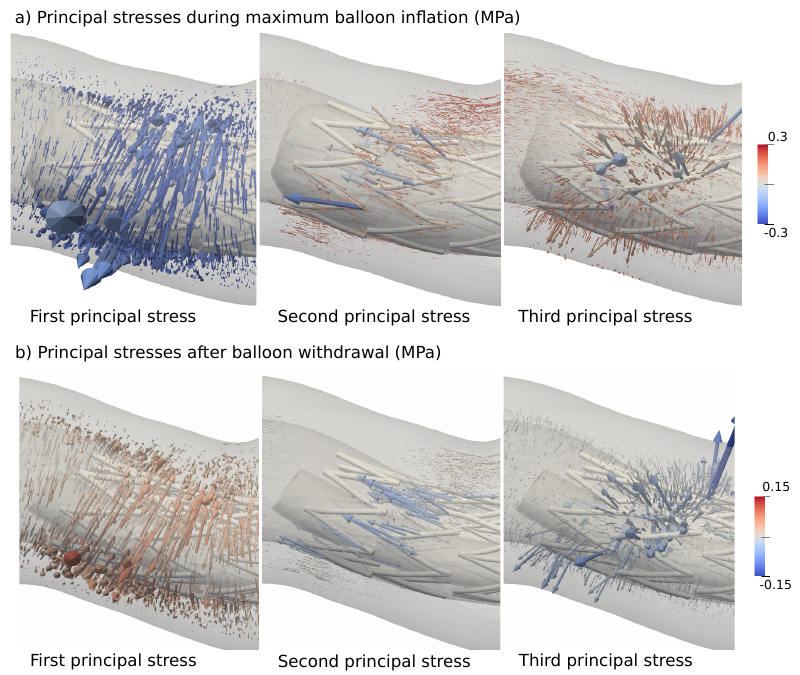}   
\caption{\revb{Principal stress directions during the maximum balloon inflation and after the balloon withdrawal.}}
\label{fig:stressdirs}
\end{figure}

\revb{Figure~\ref{fig:stressdirs} shows the directions of the principal stresses for an extract of the patient-specific numerical exemplary at the time of a) the maximum balloon inflation and b) after the balloon withdrawal.}
\revb{The arrows are scaled and colored based on the magnitude. Note that in Figure~\ref{fig:stressdirs}b), the arrow lengths are scaled with a factor of 2 compared to Figure~\ref{fig:stressdirs}a) to improve their visibility.}

\subsection{\revb{Stent bending moments}}

\begin{figure}[htbp]
\centering
\includegraphics[width=\textwidth]{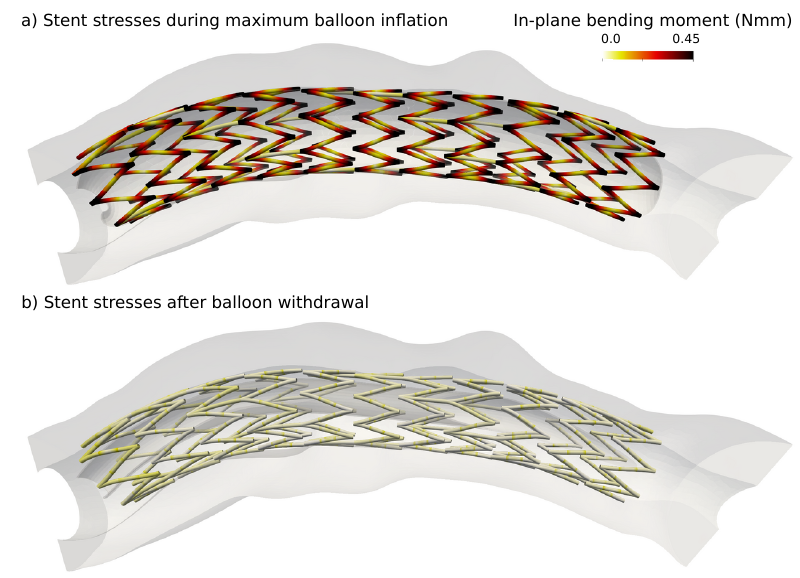}   
\caption{\revb{In-plane bending moment in the stent struts a) during and b) after PCI}}
\label{fig:stentstresses}
\end{figure}

\revb{Figure~\ref{fig:stentstresses} shows the total bending moment in the stent struts at the time of the maximum expansion and after the balloon withdrawal. 
The bending moment reaches $\unit[0.62]{Nmm}$ during the maximum balloon expansion; the bending moments are highest in the connections between the stent struts. 
After the balloon withdrawal, bending moments do not exceed $\unit[0.08]{Nmm}$ and are more evenly distributed through the stent struts.}

\subsection{\revb{Contact forces}}

\begin{figure}[htbp]
\centering
\includegraphics[width=\textwidth]{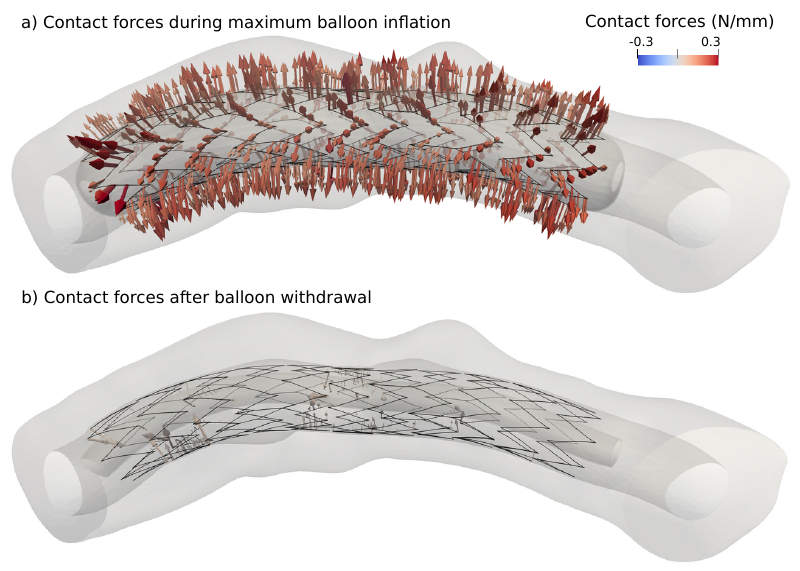}   
\caption{\revb{Contact forces between the stent and the artery a) during and b) after PCI}}
\label{fig:contactforces}
\end{figure}

\revb{Figure~\ref{fig:contactforces} shows the contact forces between the stent and the artery, i.e., line loads from the mixed-dimensional beam-to-solid interaction. 
Note that the arrow lengths in Figure~\ref{fig:contactforces}b) are scaled by a factor of 5 compared to Figure~\ref{fig:contactforces}a) to improve their visibility.
The contact forces during the balloon expansion reach $\unit[0.32]{N/mm}$ on the artery. 
The contact forces are slightly higher at the stent ends than in the center; within one stent strut, the forces are generally higher in the strut center. 
After the balloon withdrawal, contact forces from the artery to the stent reach up to $\unit[0.06]{N/mm}$ and are present in some nodes, mostly in the connections between the stent struts.}

\cleardoublepage

\bibliographystyle{elsarticle-num} 
\bibliography{bibliography3}

\end{document}